\documentclass[twocolumn,showpacs,preprintnumbers,amsmath,amssymb]{revtex4}

\usepackage{graphicx}
\usepackage{dcolumn}
\usepackage{bm}

\begin{document}

\preprint{}

\title{Quantum-dot-based optical polarization conversion}

\author{G.~V.~Astakhov$^{1,2}$}
\author{T.~Kiessling$^{1}$}
\author{A.~V.~Platonov$^{2}$}
\author{T.~Slobodskyy$^{1}$}
\author{S.~Mahapatra$^{1}$}
\author{W.~Ossau$^{1}$}
\author{G.~Schmidt$^{1}$}
\author{K.~Brunner$^{1}$}
\author{L.~W.~Molenkamp$^{1}$}

\affiliation{ $^{1}$Physikalisches Institut der Universit\"{a}t
W\"{u}rzburg, 97074 W\"{u}rzburg, Germany \\
$^{2}$A.F.Ioffe Physico-Technical Institute, Russian Academy of
Sciences, 194021 St.Petersburg, Russia }

\date{\today}

\begin{abstract}
We report circular-to-linear and linear-to-circular conversion of
optical polarization by semiconductor quantum dots. The
polarization conversion occurs under continuous wave excitation in
absence of any magnetic field. The effect originates from quantum
interference of linearly and circularly polarized photon states,
induced by the natural anisotropic shape of the self assembled
dots. The behavior can be qualitatively explained in terms of a
pseudospin formalism.
\end{abstract}


\maketitle

Quantum dots (QDs) are essentially zero-dimensional semiconductor
nanostructures that exhibit an atomic-like line spectrum in the
optical frequency range, and are therefore often referred to as
artificial atoms. Their small (nm-scale) size in combination with
their strong interaction with light has led to speculations about
possible applications of QDs in optical quantum computation
\cite{qubit,Optic_qubit}.

While most device concepts assume highly symmetric (circular)
dots, it is experimentally well established that self-assembled
semiconductor QDs often grow in a highly anisotropic manner,
reducing the point-group symmetry of a single QD to $C_{2v}$ or
still further. This natural shape anisotropy is reputed to be
unwanted, and its consequences for the physics of the system have
attracted only limited interest. In this Letter we demonstrate
novel physics that is the direct result of the low in-plane
symmetry of the QDs. We observe conversion of the polarization of
otical radiation from circular to linear (and \textit{vice versa})
mediated by QDs. The low symmetry of the dots naturally induces
quantum interference between linear and circular polarized photon
states. Time resolved experiments would result in quantum beats in
the polarization, while under the steady-state conditions that we
examine, a net conversion results. The cw effect has strong
analogies with the Hanle effect, be it that our results are all
obtained without an external magnetic field.

\begin{figure}[bp]
\includegraphics[width=.47\textwidth]{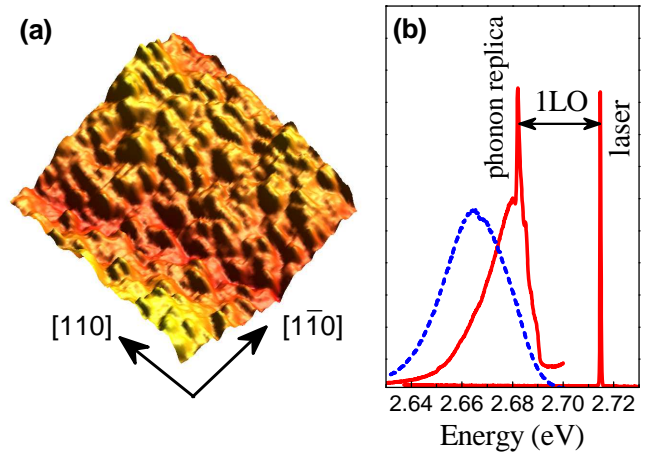}
\caption{(Color online) Characterization of CdSe/ZnSe quantum
dots. (a) Atomic force microscope image of a CdSe/ZnSe quantum dot
layer. The QDs are alongated along $\mathrm{[110]}$ axis. (b) PL
spectra for nonresonant (filled area under dotted curve) and
resonant (solid curve) excitation, respectively. The phonon
replica is well resolved in the PL spectrum as a narrow peak
separated from the laser line by the LO-phonon energy, which is
32~meV in ZnSe. } \label{fig1}
\end{figure}

The CdSe/ZnSe QDs used in our experiments are grown by molecular
beam epitaxy \cite{QDgrowth2}. A 0.3~nm thick CdSe layer is
deposited on top of a 50 nm-thick ZnSe buffer at a substrate
temperature of $300^{\circ}$C. A growth interrupt of 10 seconds
prior to capping by  25~nm ZnSe results in the formation of the
CdSe dots by self assembly. Typically, these dots are 1~nm high
and sub-10~nm in lateral dimensions, but with a high areal density
(above $10^{11}$~cm$^{-2}$). In order to image the QDs using
atomic force microscopy (AFM), also an uncapped sample has been
grown. The AFM image of this sample, presented in
Fig.~\ref{fig1}a, shows distinct elongated islands. The dots are
preferentially oriented along the $\mathrm{[110]}$ direction, in
agreement with the optical characterization discussed below. This
is quite similar to earlier studies on monolayer-fluctuation QDs
\cite{ref6}. The preferential orientation implies that the
ensemble of dots has a net spatial anisotropy which, as we will
show in the following, is essential for the polarization
conversion. The average symmetry of the ensemble of dots is
reduced to $C_{2v}$, as compared with the full $T_d$ symmetry of
the zincblende bulk lattice and the $D_{2d}$ group of the
corresponding quantum well.

\begin{figure}[tbp]
\includegraphics[width=.37\textwidth]{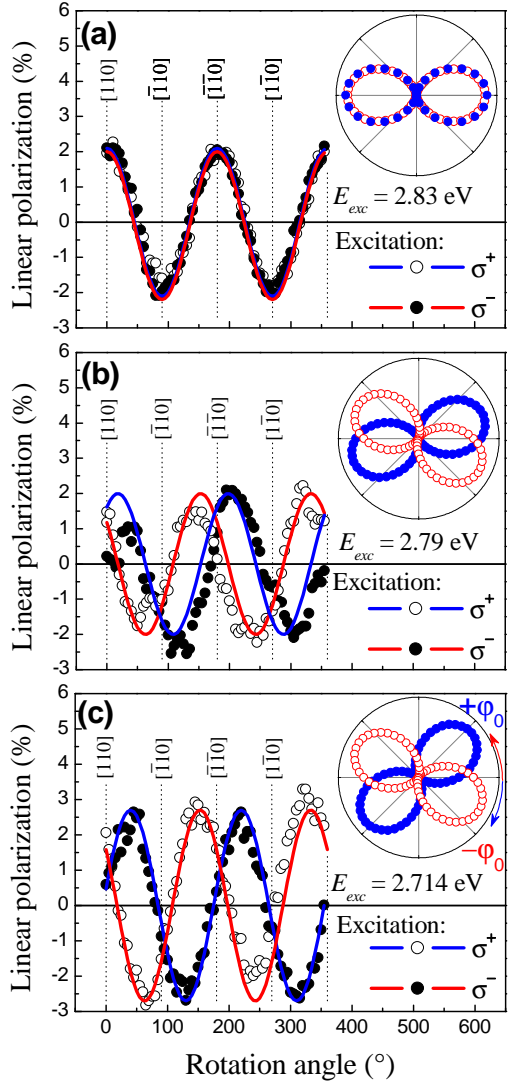}
\caption{(Color online) Circular-to-linear polarization conversion
by CdSe/ZnSe QDs. (a) Angle scans of the linear polarization
detected at the PL maximum under nonresonant excitation above the
ZnSe barrier ($E_{exc} = 2.83$~eV) with $\sigma^{+}$ (open
symbols) and $\sigma^{-}$ (solid symbols) circularly polarized
light. The solid curves are fits assuming $\rho_0 \cos 2\alpha$.
(b) Angle scans of the linear polarization detected at the PL
maximum under nonresonant excitation below the ZnSe barrier
directly in the excited states of the QDs ($E_{exc} = 2.79$~eV)
with $\sigma^{+}$ (open symbols) and $\sigma^{-}$ (solid symbols)
circularly polarized light. The solid curves are fits assuming
$\rho_0 \cos (2\alpha \mp 2\varphi_0)$, $2\varphi_0 = 44^{\circ}$.
(c) Angle scans of linear polarization detected at the phonon
replica under $\sigma^{+}$ (open symbols) and $\sigma^{-}$ (solid
symbols) circularly polarized resonant excitation ($E_{exc} =
2.714$~eV). The solid curves are again fits assuming $\rho_0 \cos
(2\alpha \mp 2\varphi_0)$, $2\varphi_0 = 67^{\circ}$. The Insets
in all panels show the same data (but shifted by a constant of
$\rho_0$ to positive values) in polar coordinates. Zero rotation
angle in all panels means that the linear analyzer is orientated
parallel to the $\mathrm{[110]}$ crystallographic direction. The
magnetic field for all data is zero.} \label{fig2}
\end{figure}

For optical excitation we use a stilbene-3 dye-laser, pumped by
the ultra-violet lines of an Ar-ion laser. For nonresonant
excitation the laser energy is tuned to $E_{exc}=2.83$~eV
(exceeding the band gap of the ZnSe barrier) or to
$E_{exc}=2.79$~eV (directly into the excited states of the QDs). A
typical photoluminescence (PL) spectrum of the QDs under
non-resonant excitation is shown in Fig.~\ref{fig1}b as the filled
area. The PL band of 30~meV width has a maximum at $E_0 =
2.665$~eV. For the angle dependent polarization data we discuss
below, the polarization is detected at the maximum of the PL band,
but we have verified the the degree of polarization does not vary
strongly over the band. Resonant excitation of the CdSe QDs is
obtained for $E_{exc}=2.714$~eV. In this case the polarization of
the PL is detected at the phonon replica, that now can be clearly
resolved in the emission spectrum (red curve in Fig.~\ref{fig1}b).

In order to investigate the in-plane optical anisotropy of the
QDs, the sample is mounted on a rotating holder. Its orientation
is controlled using a stepping motor with an accuracy better than
$1^{\circ}$. Rotation-angle dependent scans of the PL polarization
in the laboratory frame are carried out using fixed analyzers
(Glan-Thompson prisms) and a conventional optical setup consisting
of a photo-elastic modulator operating at frequency  $f = 50$~kHz
and a two-channel photon counter. The circular polarization
$\rho_{circ}^{lab}$ is detected at $f$ and the linear polarization
$\rho_{lin}^{lab}$ is detected at the double frequency $2f$. The
measured polarization degrees in the lab frame are linked to those
in the sample frame $[\rho_{l'}, \rho_{l}, \rho_{c}]$, as follows:
$\rho_{circ}^{lab}=\rho_{c}$; $\rho_{lin}^{lab}= \rho_{l'}\cos 2
\alpha \,- \rho_{l} \sin 2 \alpha$. Here $\rho_{l'} =
(I_{\mathrm{[110]}} - I_{\mathrm{[1\overline{1}0]}}) /
(I_{\mathrm{[110]}} + I_{\mathrm{[1\overline{1}0]}})$ and
$\rho_{l} = (I_{\mathrm{[100]}} - I_{\mathrm{[010]}}) /
(I_{\mathrm{[100]}} + I_{\mathrm{[010]}})$,  and $\alpha$ is the
angle between the sample and laboratory coordinate frames. For
noise reduction all optical experiments are performed at a
temperature of 1.6~K. No magnetic fields are applied.

The absorption of photons by the QDs results in the formation of
excitons, where the polarization of the photons is linked to the
spin states of the exciton. The confinement of an exciton in the
small volume of a QD leads to an enhancement of the electron-hole
exchange interaction. Due to the low symmetry of our QDs this
results in an anisotropic exchange splitting, $\hbar \Omega$.
Typically, for CdSe/ZnSe QDs, $\hbar \Omega \sim 0-0.5$~meV
\cite{ref7,ref9}. This splitting can be directly observed in the
photoluminescence spectrum of a single QD through the occurrence
of doublet emission lines. When an ensemble of QDs is probed, the
exchange splitting is buried in the much larger ($\sim 30$~meV)
inhomogeneous broadening of the PL band (Fig.~\ref{fig1}b).
However, for nonresonant excitation, above ZnSe barrier
($E_{exc}=2.83$~eV), the anisotropic exchange splitting manifests
itself as a built-in linear polarization. Figure~\ref{fig2}a shows
the degree of linear polarization measured in a fixed coordinate
basis while the sample is rotated by an angle $\alpha$. The
polarization oscillates as $\cos 2\alpha$, just as would be
observed for a linear polarizer. As can clearly be seen from the
polar plot in the inset of Fig.~\ref{fig2}a, the polarization axis
is linked to the $\mathrm{[110]}$ crystallographic direction, and
it does not depend on the handedness of the polarization of the
exciting light. This behavior is what one intuitively expects from
the shape of the QDs found in Fig.~\ref{fig1}a.

However, more counter-intuitive results are obtained under
quasi-resonant excitation ($E_{exc}=2.714$~eV). The PL spectrum of
the QDs is now dominated by a narrow peak that we attribute to a
phonon replica of the laser line (Fig.~\ref{fig1}b). It appears
due to fast excitonic recombination combined with the emission of
an LO-phonon. Under these conditions the polarization axis is no
longer fixed to the $\mathrm{[110]}$ crystalline direction. As
shown in Fig.~\ref{fig2}c, the angle dependence of the linear
polarization now varies as $ \cos (2\alpha \mp 2\varphi_0)$, where
the sign depends on the handedness of the circularly polarized
excitation light and $2\varphi_0 = 67^{\circ}$. This behavior is
ever so more clearly apparent from the polar plot in the inset of
Fig.~\ref{fig2}c. The polarization axis is rotated away from
$\mathrm{[110]}$ by an angle $\varphi_0$, counter-clockwise
towards the $\mathrm{[010]}$ direction for $\sigma^{+}$, and
clockwise towards the $\mathrm{[100]}$ direction for $\sigma^{-}$
polarization of the incoming light. Such a behavior implies,
indeed, circular-to-linear polarization conversion.

In order to estimate the conversion efficiency under
$\sigma^{\pm}$ circular-polarized excitation, denoted by $P_c =
\pm 1$, we describe the total polarization of the emitted light by
a vector $[\rho_{l'}, \rho_{l}, \rho_{c}]$ inside a Poincar\'{e}
sphere defining a novel type of quasi-spin, or two-level system.
Here, $\rho_{l'}$ is the linear polarization along
$\mathrm{[110]}$, $\rho_{l}$ is the linear polarization along
$\mathrm{[100]}$, and $\rho_{c}$ is the circular polarization.
These Stokes coordinates satisfy $\sqrt{{\rho_{l'}}^2 +
{\rho_{l}}^2 +{\rho_{c}}^2} \leq 1$. As efficient conversion we
define the condition $\rho_{l} > \rho_{l'}$ and $\rho_{l} >
\rho_{c}$. According to Fig.~\ref{fig2}c the maximum amplitude of
the linear polarization is  $\rho_{0} = \sqrt{{\rho_{l'}}^2 +
{\rho_{l}}^2} = 2.7 \%$, so we have $\rho_{l} = \rho_{0} \sin
2\varphi_0 = 2.5 \%$ and $\rho_{l'} = \rho_{0} \cos 2\varphi_0 =
1.0 \%$. We have also measured the optical orientation
\cite{ref11}, i.e. the degree of circular polarization of the
emitted light under circularly polarized excitation and obtained
$\rho_{c} \approx 1 \%$. For the experimental values the above
condition of efficient conversion is obviously fulfilled.

Polarization conversion in low dimensional systems has been
predicted by Ivchenko \textit{et al.} \cite{ref12}. In the
presence of a preferential direction for the excitonic states in
QDs, the circularly and linearly polarized contributions to the
emission can show quantum interference (e.g., quantum beats in the
time domain). Obviously, an external magnetic field can induce
this preferential direction. Meanwhile, magnetic field-induced
polarization conversion has been demonstrated experimentally in
superlattices \cite{ref13} and QDs \cite{ref13b}. However, using
the anisotropic exchange interaction to define the preferential
direction induces a beating of the circular and the
$\mathrm{[100]}$ linear polarizations even in zero magnetic field.
In the simplest case the time evolution after circularly polarized
excitation $P_c$ at $t=0$ can be expressed as $\rho_{c}(t) = P_c
\cos (\Omega t) \exp (-t/\tau_s)$ and $\rho_{l}(t) = P_c \sin
(\Omega t) \exp (-t/\tau_s) $. The circular and linear
polarizations thus beat in antiphase, decaying with spin coherence
time $\tau_s$ to zero. This has been partly verified previously in
quantum beat experiments \cite{ref14, ref14a} where precession of
the linear (circular) polarization component excited with linearly
(circularly) polarized light at Larmor frequency $\Omega$ was
observed.

Within the pseudospin formalism \cite{ref13,ref13a} the Stokes
coordinates in the Poincar\'{e} sphere are linked to a pseudospin
$\mathbf{S}$ by the simple relation
\begin{equation}
\rho_{l'} = S_1 \,, \,\,\, \rho_{l} = S_2 \,, \,\,\, \rho_{c} =
S_3 \,. \label{eq1}
\end{equation}
The $S_1/2$, $S_2/2$, and $S_3/2$ behave as $x$-, $y$-, and
$z$-projections of a spin in real space. In zero magnetic field
the pseudospin Hamiltonian can be written in the form
\begin{equation}
\mathcal{H} = \frac{\hbar}{2}\Omega \sigma _x \,, \label{eq2}
\end{equation}
where $\sigma _x$ is the Pauli matrix. The dynamics of the
polarization of the PL described by the vector $\mathbf{S}$ after
$\mathbf{P_{ex}}$-polarized excitation is given by \cite{ref11}:
\begin{equation}
\frac{\partial \mathbf{S}}{\partial t} = \mathbf{\Omega} \times
\mathbf{S} - \frac{\mathbf{S} - \mathbf{P_{eq}}}{\tau_s} -
\frac{\mathbf{S} - \mathbf{P_{ex}}}{\tau_0}\,. \label{eq2a}
\end{equation}
Here $\tau_0$ is the exciton life time and $\mathbf{P_{eq}}$ is
equilibrium polarization of the emission. According to our
Hamiltonian~(\ref{eq2}) $\mathbf{\Omega} = [\Omega,0,0]$, and
$\mathbf{P_{eq}} = [\Upsilon_{lin},0,0]$ where the built-in linear
polarization $\Upsilon_{lin}$ originates from the linear dichroism
of the QDs and thermal population of the exchange-split states.
Eq.~(\ref{eq2a}) can be solved for steady-state conditions (i.e.,
under cw excitation) when the PL is excited by circularly
polarized light $\mathbf{P_{ex}} = [0,0,P_c]$, yielding
\begin{eqnarray}
\rho_{l'} = \frac{T}{\tau _s} \Upsilon_{lin} \,\,\,\,\,\,\,\,\,\,\,\,\,\,\,\,\,\,\,\,\,\,\,\,\,\, \nonumber \\
\rho_{l} = -\frac{T}{\tau_0} \frac{\Omega T}{1+(\Omega T)^2} P_c \,\,\, \nonumber \\
\rho_{c} = \frac{T}{\tau_0} \frac{1}{1+(\Omega T)^2} P_c \,\,,
\label{eq3}
\end{eqnarray}
where $T^{-1} = {\tau_s}^{-1} + {\tau_0}^{-1}$. The second
identity describes, indeed, circular-to-linear polarization
conversion. We note that the QD ensemble is inhomogeneous, i.e.,
the anisotropic exchange splitting varies from dot to dot. This
can be taken into account by using average values  $\langle \Omega
\rangle$, $\langle \Omega^2 \rangle$.

Equations~(\ref{eq3}) successfully explain the polarization
behavior presented in Fig.~\ref{fig2}. In the case of
quasi-resonant excitation the PL life time is nearly equal to
radiative recombination time of the exciton $\tau_0 \simeq
\tau_r$, as $\tau_r \sim 300$~ps \cite{ref17}. According
Eqs.~(\ref{eq3}), the observation of efficient conversion in
Fig.~\ref{fig2}c implies $\tau_0 \leq \tau_s$. This condition is
in agreement with the generally expected long spin coherence time
in the QD ground state. E.g., the spin relaxation time of a single
hole was found to be about 10~ns \cite{ref15}. In the case of
nonresonant excitation below the ZnSe barrier into the excited
states of the CdSe dots ($E_{exc}=2.79$~eV), $\tau_0$ is also
partly determined by the relaxation into the QD ground state.
During this process the spin coherence is partially lost, as is
also independently confirmed by experiments on energy-dependent
optical orientation \cite{LO1,LO2}. As a result, the conversion in
Fig.~\ref{fig2}b is not optimal. We find $2\varphi_0 =
44^{\circ}$, which corresponds $\rho_{l'} \approx \rho_{l} \approx
1.4 \%$. This implies that the condition for efficient conversion
is no longer satisfied. Upon excitation above the ZnSe barrier
($E_{exc}=2.83$~eV) an electron and a hole are trapped by a QD
independently, so they do not form a coherent spin states. As a
consequence, Fig.~\ref{fig2}a shows no conversion at all,
$2\varphi_0 = 0^{\circ}$.

Eqs. ~(\ref{eq3}) are simple but essential for the QD conversion
mechanism. The third identity in Eqs.~(\ref{eq3}) is very similar
to the Hanle effect, with the Zeeman splitting induced by a
magnetic field replaced by the zero-field anisotropic exchange
splitting. In quantum dots the anisotropic exchange splitting
$\hbar \Omega$ is an order of magnitude larger than in
superlattices \cite{ref13}. As a result the polarization
conversion under cw excitation is significant. The conversation
factor is  $K = \rho_{l} / \rho_{c} = \langle \Omega \rangle T$.
In QDs $\Omega T$ is typically in the range of $0-100$, which is
in good agreement with the present experimental data, as we found
$K \approx 3$. It also follows from equations~(\ref{eq3}) that for
$\Omega T =1$ and $\tau_0 \ll \tau_s$ the polarization reaches
$\rho_{c}=\rho_{l}=50\%$.

\begin{figure}[tbp]
\includegraphics[width=.37\textwidth]{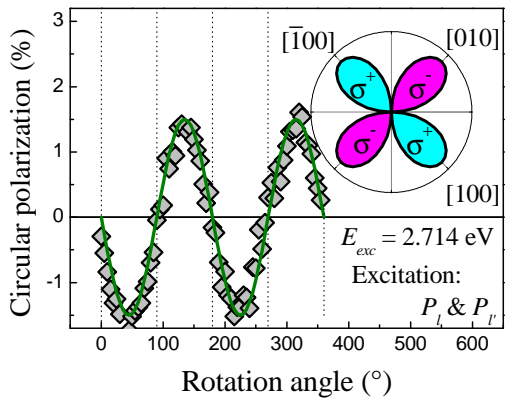}
\caption{(Color online) Linear-to-circular polarization conversion
by CdSe/ZnSe QDs. It reveals itself in an angle scan of circular
polarization detected at the phonon replica under linearly
polarized resonant excitation ($E_{exc} = 2.714$~eV). The curve is
a fit, assuming $\rho_0 \sin 2\alpha$. The Inset shows absolute
value of the same data $|\rho_0 \sin 2\alpha|$ in polar
coordinates. Zero rotation angle means that the linear polarizer
is orientated parallel to the $\mathrm{[110]}$ crystallographic
direction. The magnetic field is zero. } \label{fig3}
\end{figure}

The most intriguing effect is the counter-conversion, i.e.,
conversion from linear to circular polarization, which can occur
for our dots due to time reversal symmetry. Indeed we observe this
effect, as shown in Fig.~\ref{fig3}. With linear polarized
excitation along $\mathrm{[100]}$, $\sigma^{+}$ polarized emission
appears. The effect changes sign to $\sigma^{-}$ when excited
along $\mathrm{[010]}$. No conversion is observed when the linear
polarizer at the excitation was oriented along $\mathrm{[110]}$ or
$\mathrm{[1\overline{1}0]}$ directions. This behavior is in a good
qualitative agreement with theory, and obeys similar equations as
Eqs.~(\ref{eq3}) upon interchange of the indices $l
\leftrightarrow c$ and reversing the sign in the second identity.

All experiments discussed above were obtained for quantum dots
containing no electrons. In negatively charged QDs, containing a
single extra electron, the anisotropic exchange splitting is
modified drastically. With a photo-created electron the extra
electron forms the energetically favorable singlet state with zero
total electron spin. Since the electron-hole exchange interaction
is proportional to the spins \cite{ref10} of electrons and holes,
the anisotropic exchange splitting in a charged QD equals exactly
zero ($\hbar \Omega = 0$). By applying a bias voltage, additional
electrons can be pushed into or out of the QDs \cite{QD_voltage}.
This may provide extra functionality to the QD converter, and may
provide a flexible approach for spin-based electro-optical
devices. Due to the optical selection rules \cite{ref11}, the spin
of a photo-excited electron in the conduction band is proportional
to the photon's circular polarization. Thus, instead of directly
manipulating electron spin one can alternatively control the light
polarization within the same circuit.

In summary, we have demonstrated efficient circular-to-linear and
linear-to-circular light polarization conversion by quantum dots.
The conversion occurs in zero magnetic field and is induced by
anisotropic exchange splitting. For optimized QD dimensions
conversion efficiencies up to $50\%$ can be achieved. An important
advantage of the QD converter is the possibility of control of the
optical activity by charging the dots by application of a bias
voltage. Our findings may have obvious practical applications in
information processing as the dots can easily be integrated in
semiconductor circuits.

The authors thank V.~L.~Korenev for fruitful discussions. This
work was supported  by the Deutsche Forschungsgemeinschaft (SFB
410) and RFBR.

\end{document}